\documentclass[iop]{emulateapj}

\slugcomment{Accepted for publication in the Astrophysical Journal}

\shorttitle{A Spectroscopic Binary in the Hercules dwarf spheroidal}
\shortauthors{A. Koch et al.}

\begin{document}

\title{A spectroscopic binary in the Hercules dwarf spheroidal galaxy\altaffilmark{$\dagger$}}

\author{
Andreas Koch\altaffilmark{1},   
 Terese Hansen\altaffilmark{1},  
 Sofia Feltzing\altaffilmark{2}, 
\& Mark I. Wilkinson\altaffilmark{3}
}
\altaffiltext{$\dagger$}{This paper includes data gathered with the 6.5 meter Magellan Telescopes located at Las Campanas Observatory, Chile, 
and is in parts based on observations made with ESO Telescopes at the  Paranal Observatory under programmes ID 079.B-0447(A) and 083.D-0688(A).}
\altaffiltext{1}{Landessternwarte, Zentrum f\"ur Astronomie der Universit\"at Heidelberg, K\"onigstuhl 12, 69117 Heidelberg, Germany}
\altaffiltext{2}{Lund Observatory, Department of Astronomy and Theoretical Physics, Box 43, SE-221 00, Lund, Sweden}
\altaffiltext{3}{Department of Physics and Astronomy, University of Leicester, University Road, Leicester LE1 7RH}
\email{akoch@lsw.uni-heidelberg.de}
\begin{abstract}
We present the radial velocity curve of a single-lined spectroscopic binary in the faint Hercules dwarf
spheroidal (dSph)  galaxy, based on 34 individual spectra covering more than two years of observations. This is the first time that  
orbital elements could be derived for a binary in a dSph. The system consists of a metal-poor red giant and  a low-mass companion, possibly a white dwarf, with  
a 135-days period in a moderately eccentric ($e=0.18$) orbit.
{ Its period and eccentricity are fully consistent with metal-poor binaries in the Galactic halo, 
while the projected  semimajor axis is small, at  $a_p\,\sin\,i = 38$ $R_{\sun}$. In fact, 
a very  close orbit} 
could inhibit the production of heavier elements through $s$-process nucleosynthesis, leading to the 
very low abundances of neutron-capture elements that are found in this star.  
We discuss the further implications for the chemical enrichment history of the Hercules dSph, { but find no compelling binary scenario that could reasonably explain the 
full, peculiar abundance pattern of the Hercules dSph galaxy.
}
\end{abstract}
%
%
%
%
\section{Introduction}
Binaries in old stellar populations play an important role in setting the chemical abundance patterns observed in 
the oldest stars. 
For instance, mass transfer from an Asymptotic Giant Branch (AGB) companion can explain many of the 
$s$-process and light-element distributions seen in a variety of  meta- poor stars in the 
Galactic halo (e.g., Masseron et al. 2010). 

Often unknown binary fractions prevent reliable mass  estimates of the dark matter content 
of  low-luminosity dwarf spheroidal (dSph) galaxies, since the binaries can inflate and thus falsify their measured velocity dispersions  
(e.g., Hargreaves et al. 1996; Koch et al. 2007a; McConnachie et al. 2010). 

So far,  data on binaries in the outer halo,  globular clusters, and dSphs are sparse, since the stars are often faint so that  multi-epoch radial velocity 
 measurements are expensive. Repeat measurements of stellar kinematics, sometimes by different groups,  often reveal ``discrepancies'' between 
measurements of a given star, which is then taken as evidence of ``binarity''. 

Here, we present velocity variations in a red giant of the faint Hercules dSph (Belokurov et al. 2007),  
Her-3 (Koch et al. 2008; this is \#41082 in Ad\'en et al. 2009; 2011). 
Previous studies have identified it with a metal-poor ([Fe/H]$=-2.04$ dex) member of the dSph, and it lies towards the more metal-rich tail of Hercules 
owing to the galaxy's very extended metallicity distribution (reaching below $-3$ dex; e.g., Ad\'en et al. 2009, 2011).  
One of the most striking features of Hercules is a, seemingly, overall deficiency in neutron-capture elements (represented by the Ba-measurements of Koch et al. 2013), 
along with abundance patterns in individual stars including Her-3 that suggest that the galaxy was imprinted with the ejecta of only a few, massive type II supernovae (SNe II). 
In the present work we attempt to connect the chemical abundance anomalies with the discovery of Her-3's binarity. 
\section{Spectroscopic data}
Figure~1 shows a chart of the surroundings of Her-3.
\begin{figure}[!ht]
\begin{center}
\includegraphics[angle=0,width=1\hsize]{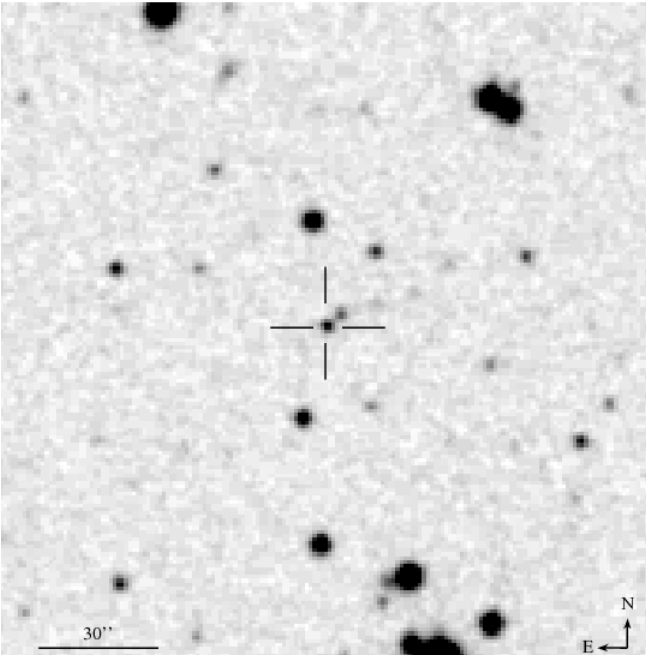}
\end{center}
\caption{R-band image of Her-3 from the DSS. The Image covers $1.5\arcmin\times1.5\arcmin$. Note that the object NW to Her-3 is not the secondary (Sect.~5).}
\end{figure}
This star was first identified as a possible binary when combining the radial velocity data  of Koch et al. (2008) with those of Ad\'en et al. (2009), where we noted a systematic offset that 
could not be explained by differences in the instrumental set-up, reduction techniques, nor by the measurement uncertainties.
Subsequent data were then obtained and homogeneously analysed to construct the final radial velocity curve used in this work. 
In the following we briefly recapitulate the details of the individual observing runs. 
\subsection{MIKE}
The first set of observations was carried out over two nights in July 2007 
using the Magellan Inamori Kyocera Echelle (MIKE) 
spectrograph at the 6.5-m Clay Telescope; these consisted of seven individual exposures of 45--60 min each, which 
were ultimately combined into the spectrum used for abundance analysis in Koch et al. (2008). 
The median seeing was 0.8$\arcsec$, resulting in a signal-to-noise  ratio (SNR) of the combined spectrum of 32 pixel$^{-1}$ at 6500\AA. 
Individual exposures are subject to lower SNRs of $\sim$12 pixel$^{-1}$, but still allowed us to perform precise radial velocity measurements (Sect.~3). 
At the chosen slit width (1$\arcsec$) and on-chip binning (2$\times$2) in spectral and spatial direction we could obtain a resolution of $R\sim20000$. 

Subsequently, five more exposures were obtained with MIKE in Apr -- Jul 2008. Varying observing conditions prompted the use of different slit widths 
(0.5$\arcsec$ and 0.7$\arcsec$) and binnings (1$\times$1, 2$\times$2)
so that also the resolution varied, reaching as high as 40000. These data were reduced (Kelson et al. 2003) and analysed identically to the ones mentioned above. 
 \subsection{FLAMES}
The Hercules dSph was also targeted within two observational campaigns with the multiobject Fibre Large Array Multi Element Spectrograph (FLAMES) at the Very Large Telescope,   
both of which included spectra of Her-3. 
These were published by  Ad\'en et al. (2009, 2011) and we refer the reader to those works for details on the observation, reduction, and analysis strategies.
Suffice it to say that the first set (Ad\'en et al. 2009) consisted of seven spectra taken with grating LR8, 
which covers the prominent near-infrared calcium triplet (CaT) at 
$\lambda\lambda$8498, 8542, 8664\AA, albeit at lower resolutions of $R\sim8500$. 
Secondly, we employed the higher-resolution ($R\sim20000$) spectra of Ad\'en et al. (2011), which comprise 17 exposures that used grating HR13, thereby covering the wavelength range of 
6100--6400\AA~(see also Koch et al. 2013). 
While not used for abundance analysis in Ad\'en et al. (2011), their final, co-added spectrum was of comparable quality to that of Koch et al. (2008). 
\section{Radial velocity measurements} 
Individual velocity measurements of Her-3 were based on cross-correlation of the spectra using IRAF's {\em fxcor} task. 
One concern is the choice of a proper template, as a target-template mismatch can yield falsified velocities ({e.g., Gullberg \& Lindegren 2002) or lead to unnecessarily 
weak peaks in the cross-correlation function. 
Fortunately, the stellar parameters and chemical composition of the red giant primary are well known (Koch et al. 2008). Thus 
we computed a synthetic spectrum with those parameters over the maximum wavelength region 
covered by the observations (i.e., 4500--9000\AA) to serve as a template in the cross-correlation of the higher-resolution, optical 
spectra. For the LR8 spectra around the CaT we used a simple Gaussian model of the three strong CaT lines. This procedure
has proven to yield stable results over a broad range of stellar types  (e.g., Kleyna et al. 2004). 
Table~2 lists the resulting velocities, each corrected to the heliocentric rest frame. 
\begin{center}
\begin{deluxetable}{ccccc}
\tabletypesize{\scriptsize}
\tablecaption{Radial velocity curve. Times are given at mid-exposure.}
\tablewidth{0pt}
\tablehead{\colhead{HJD} & \colhead{v$_{\rm HC}$}  & \colhead{$\sigma$v$_{\rm HC}$}  & \colhead{} & \colhead{}\\
($-$2454000) & \colhead{[km\,s$^{-1}$] }  &\colhead{[km\,s$^{-1}$] }  & \raisebox{1.5ex}[-1.5ex]{ Phase\tablenotemark{a}} & \raisebox{1.5ex}[-1.5ex]{Instrument}
}
\startdata
%
%
 205.8004 & 51.94 &  1.19  & 0.3587  &  FLAMES, LR8 \\
 205.8391 & 55.96 &  0.72  & 0.3590  &  FLAMES, LR8 \\
 230.6579 & 40.89 &  4.71  & 0.5425  &  FLAMES, LR8 \\
 237.7139 & 46.51 &  0.95  & 0.5946  &  FLAMES, LR8 \\
 272.5700 & 46.08 &  1.41  & 0.8523  &  FLAMES, LR8 \\
 272.6015 & 46.16 &  1.24  & 0.8525  &  FLAMES, LR8 \\
 272.6365 & 48.87 &  1.16  & 0.8528  &  FLAMES, LR8 \\
 292.4781 & 65.76 &  0.74  & 0.9995  &  MIKE \\
 292.6510 & 65.48 &  0.72  & 0.0008  &  MIKE \\
 292.6934 & 66.58 &  0.96  & 0.0011  &  MIKE \\
 293.4926 & 66.63 &  0.66  & 0.0070  &  MIKE \\
 293.5349 & 66.06 &  0.85  & 0.0073  &  MIKE \\
 293.5771 & 66.36 &  0.77  & 0.0076  &  MIKE \\
 293.6194 & 65.92 &  0.67  & 0.0079  &  MIKE \\
 581.8018 & 77.10 &  1.66  & 0.1382  &  MIKE \\
 631.6354 & 40.39 &  2.13  & 0.5066  &  MIKE \\
 631.6504 & 41.02 &  2.13  & 0.5067  &  MIKE \\
 969.6730 & 66.93 &  2.17  & 0.0054  &  FLAMES, HR13 \\
 969.7198 & 67.57 &  0.65  & 0.0057  &  FLAMES, HR13 \\
 969.7670 & 67.00 &  0.70  & 0.0061  &  FLAMES, HR13\\
 972.6734 & 67.11 &  0.98  & 0.0275  &  FLAMES, HR13 \\
 972.7221 & 66.17 &  0.54  & 0.0279  &  FLAMES, HR13 \\
 974.6697 & 68.75 &  0.74  & 0.0423  &  FLAMES, HR13 \\
 974.7185 & 71.44 &  1.98  & 0.0427  &  FLAMES, HR13 \\
 975.6531 & 68.18 &  1.76  & 0.0496  &  FLAMES, HR13 \\
 975.7369 & 70.23 &  1.63  & 0.0502  &  FLAMES, HR13 \\
 976.7332 & 69.18 &  2.73  & 0.0575  &  FLAMES, HR13 \\
 977.6417 & 72.79 &  0.96  & 0.0643  &  FLAMES, HR13 \\
 977.6896 & 65.11 &  1.48  & 0.0646  & FLAMES,  HR13 \\
 977.7366 & 69.06 &  0.59  & 0.0650  &  FLAMES, HR13 \\
 978.7481 & 69.30 &  1.96  & 0.0724  &  FLAMES, HR13 \\
 996.6116 & 68.74 &  0.54  & 0.2045  &  FLAMES, HR13 \\
 996.6590 & 61.05 &  0.82  & 0.2048  & FLAMES,  HR13 \\
1001.6200 & 66.17 &  0.89  & 0.2415  &  FLAMES, HR13 
\enddata
\tablenotetext{a}{Using the best-fit orbital solution, see Table~2.}
\end{deluxetable}
\end{center}
\section{Orbital solution}
Fig.~2 shows the radial velocity data, phased with respect to the best orbital solution that was obtained as outlined below. 
\begin{figure}[!ht]
\begin{center}
\includegraphics[angle=0,width=1\hsize]{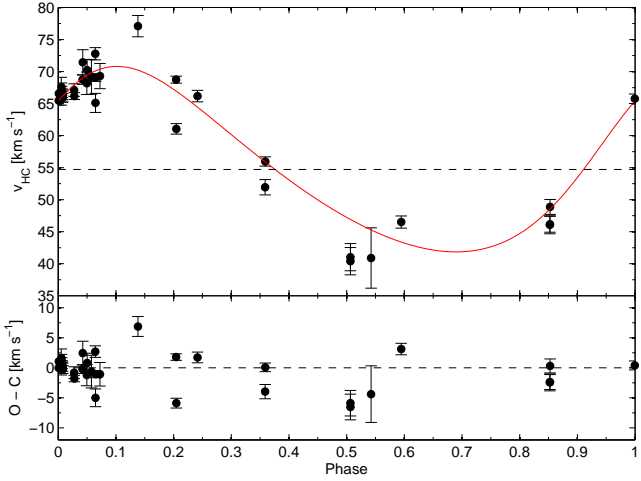}
\end{center}
\caption{
Radial velocity curve for the best orbital solution of Her-3 
 with all individual, observed velocities. 
 The lower panel shows the  $\mathcal{O-C}$ residuals with respect to our solution.}
\end{figure}

An initial solution for the orbital elements was obtained using 
  a fitting routine of a Keplerian velocity curve  (Buchhave et al. 2010), based on the formalism of P\'al
  (2009). 
   Thus, an initial orbit with a period of P=134.7 days and a semi-amplitude of K=15.2 km\,s$^{-1}$ were found. 
Starting from these values, the elements were refined by an  error-weighted Markov Chain Monte Carlo (MCMC) 
approach,  using the   Metropolis-Hastings algorithm. 
To test the stability of the most likely solutions, we 
ran 50 chains of 10$^5$ trials each. 
As the best set (Table~2) we adopted the median of the posteriori likelihood distributions for each parameter and 
the uncertainties stated in the Table are their 1$\sigma$ intervals. 
The resulting mean velocity, $\gamma$, of Her-3 is fully  consistent with the systemic mean of the Hercules dSph (Simon \& Geha 2007; Ad\'en et al. 2009).
\begin{table}[htb]
\caption{Orbital parameters}             
\centering          
\begin{tabular}{lr}     
\hline\hline       
Parameter & Value\\
\hline 
\multicolumn{2}{c}{\em Measured}\\
$P$ [days] & 135.28\,$\pm$\,0.33 \\
$T$ [HJD] & 24545563.11\,$\pm$\,7.44 \\
$\gamma$ [km\,s$^{-1}$] & 54.72\,$\pm$\,0.55 \\
$K_p$ [km\,s$^{-1}$] & 14.48\,$\pm$\,0.82 \\
$e$ & 0.18\,$\pm$\,0.06 \\
$\omega$ [deg] & 308.87\,$\pm$\,22.53 \\
\hline
\multicolumn{2}{c}{\em Derived}\\
$a_p\,\sin\,i$ [$R_{\sun}$] & 38.12\,$\pm$\,2.21 \\
$f(M)\,\sin^3\,i$ [M$_{\sun}$] & 0.0406\,$\pm$\,0.0053 \\
\hline                  
\end{tabular}
\end{table}

All of the orbital elements from the 50 MCMC chains agree with each other:
the 1$\sigma$ dispersion between the 50 runs is typically much smaller  than 
the mean error from each of the 10$^5$ MCMC trials, thus   rendering our best solution 
stable and reproducable. 
Overall, the root-mean-square of the residuals ($\mathcal{O-C}$; bottom panel of Fig.~2) is 2.8 km\,s$^{-1}$ and we find a  
reduced $\chi^2$ of 6.6. 

With a moderate eccentricity of $e=0.18$, the Her-3 system follows the trend of typical intermediate-period binaries, which 
show a wide range in $e$ in contrast to the mainly circular orbits of systems with $P<20$ days (Duquennoy \& Mayor 1991; 
Latham et al. 2002).   
While tidal circularization on the main sequence is an important asset for the oldest binaries in the Galaxy, 
this was apparently not effective in Her-3, although we can assume that it is an old ($\sim12$ Gyr) system\footnote{ 
Likewise, the very low metallicity of Her-3, at [Fe/H]=$-2.04$ dex (Koch et al. 2008) is unlikely to have any influence on the orbital properties; 
using Galactic disk and halo stars, Latham et al. (2002) showed that metallicity has only a small effect on fragmentation that leads to the binary formation.}. 
\section{Musings on the secondary}
The closest object to Her-3, still  visible on optical images (Fig.~1),  lies at a projected separation of  $\sim5\arcsec$. 
This star is fainter\footnote{The fainter star's $g$,$r$, and $i$ of (21.0, 19.6, 18.9) mag 
from the SDSS (Adelman-McCarthy et al. 2006) compare to magnitudes of (19.6, 18.6, 18.2) mag for Her-3. Also the Str\"omgren photometry of Ad\'en et al. (2009) identify it as a 
redder foreground dwarf.}
 than Her-3 
and will only contribute negligible flux ($\sim10^{-8}$) to spectroscopic measurements of the primary. 
At the distance of Hercules ($\sim$140 kpc;  Belokurov et al. 2007), this would correspond to a true separation of 3.4 pc or 700,000 AU. 
We can compare this value to the dimensions of the widest, known binaries in the Milky Way, 
where, e.g.,  Quinn et al. (2009) confirmed binarity of  common proper motion pairs in the Galactic halo as wide as 1.1 pc. 
It is unlikely that Hercules hosts binaries at such wide separations unless its low-density environment would favor their formation
(e.g., Duquennoy \& Mayor 1991; Duch\^ene \& Kraus 2013); moreover, the 
ensuing orbital periods would be considerably longer than the 135 days of Her-3.

The mass function, $M_s ^3/ (M_s + M_p)^2\,\sin^3 i$, we derived from the orbital parameters of the system is 0.04 M$_{\odot}$. 
Under certain assumptions,  this can give us an estimate of the secondary star's properties. 
Since Hercules is an old, metal-poor system (Belokurov et al. 2007; Ad\'en et al. 2009), we compared the stellar parameters of Her-3 with 
a Dartmouth isochrone (Dotter et al. 2008) of 12 Gyr and [Fe/H]=$-2$, which implied a stellar mass for 
the primary of M$_p=0.81$ M$_{\odot}$.  
Under the conservative assumption of an orbital inclination of the system of $i=90^{\circ}$, 
the mass function yields a lower limit on the secondary mass\footnote{ The resulting mass ratio of $q>0.47$ pushes Her-3 towards the second, higher-$q$ peak 
of the halo distribution identified by Goldberg et al. (2003).} 
of  M$_s>0.38\pm0.02$ M$_{\odot}$. 
The same isochrone would imply a radius of the secondary of $\sim$0.35 R$_{\odot}$ and render it a K5V main sequence dwarf. 
However, with no photometric information on the secondary it is equally likely to be a white dwarf (e.g., De Gennaro et al. 2008).  
At the implied magnitude of V$\sim$30 mag chances of direct detection are essentially null, which is confirmed by the lack of any obvious variability 
in its time-series photometry (I. Musella, priv. comm.).

More critical is the projected semi major axis of the primary that, at $a_p\,\sin\,i = 38$ R$_{\odot}$  is of similar order of magnitude as 
the primary stellar radius ($\sim$55 R$_{\odot}$  from the same isochrones). 
While this suggests the possibility that Her-3 is in a close binary with the secondary white dwarf currently sharing parts of the red giant primary's 
envelope, the small separation is very likely an inclination effect: Setting the constraint that the separation must not exceed the giant's Roche lobe (Eggleton 1983)
would translate into an inclination of $\sim18\degr$. 
\section{ Comparison with halo binaries}
{ The distribution of binary orbital elements in dwarf galaxies is almost unconstrained. While some attempts have been made to 
test the consistency of the distributions with those in the Solar neighbourhood (e.g. Minor 2013), direct determinations of the element
distributions are still lacking. Instead, we now place Her-3 in the context of typical binaries in the Milky Way Milky Way, in particular the metal-poor halo.

Based on the extensive binary sample of Carney et al. (1994), Latham (2004) noted only subtle differences 
in the characteristics, viz. period, eccentricity and fraction, of binaries in the Solar neighbourhood (Duquennoy \& Mayor 1991) and metal-poor giants in the halo field and globular clusters
(see also Carney et al. 2003; Zapatero-Osorio et al. 2004). 
In particular, the binary fraction, $f_b$, was found to be the same for the Solar-metallicity main sequence stars and metal-poor RGB stars, indicating 
that the status of binarity is not affected during the course of stellar evolution. 
However, considerable differences in the semimajor axis distributions of Population I and II stars were found by Zinnecker et al. (2004).

In Fig.~3, we show the eccentricity-period relation for halo stars from  Latham et al. (2002) and overplot our measurements for Her-3.  
\begin{figure}[!ht]
\begin{center}
\includegraphics[angle=0,width=1\hsize]{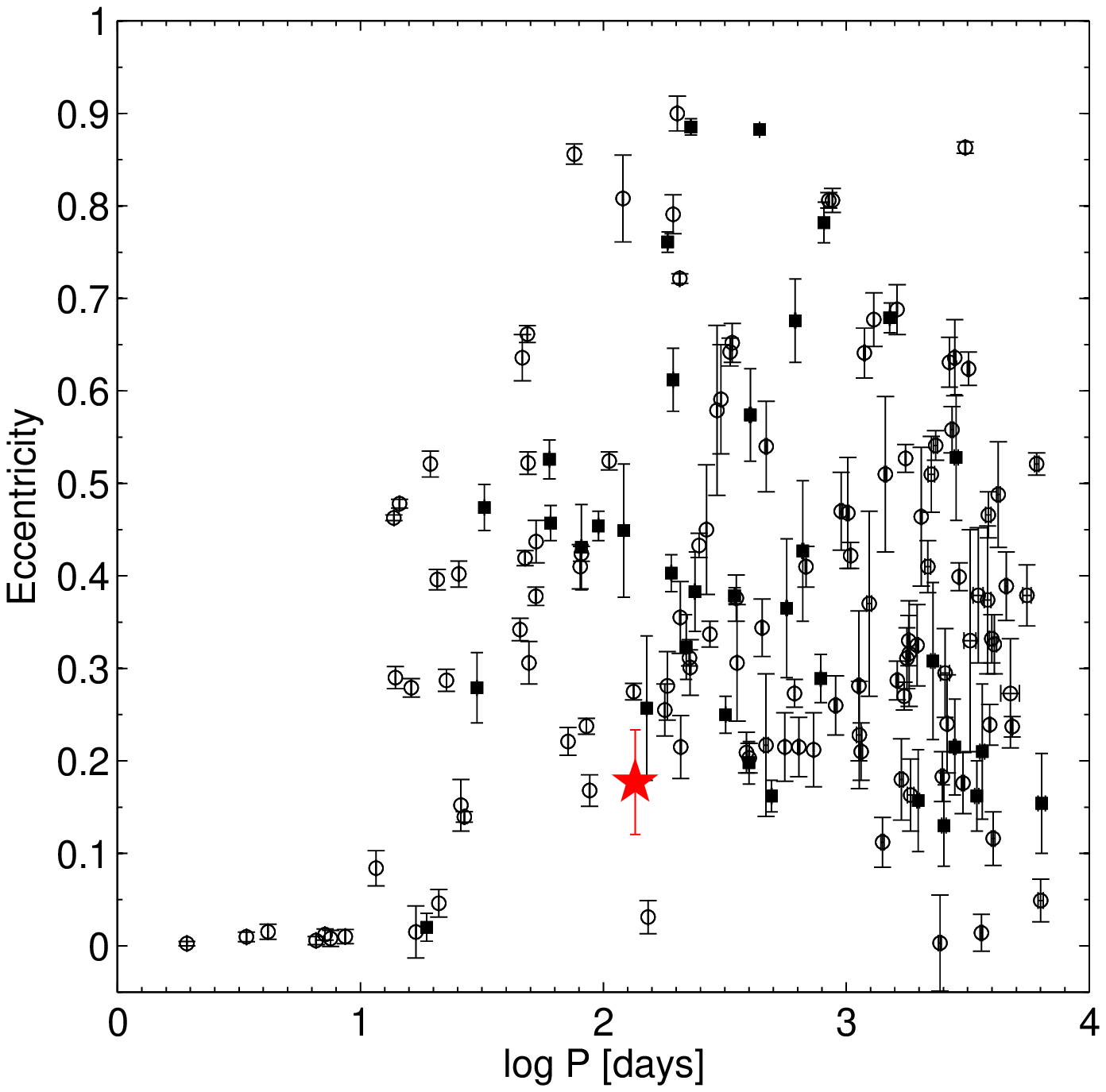}
\includegraphics[angle=0,width=1\hsize]{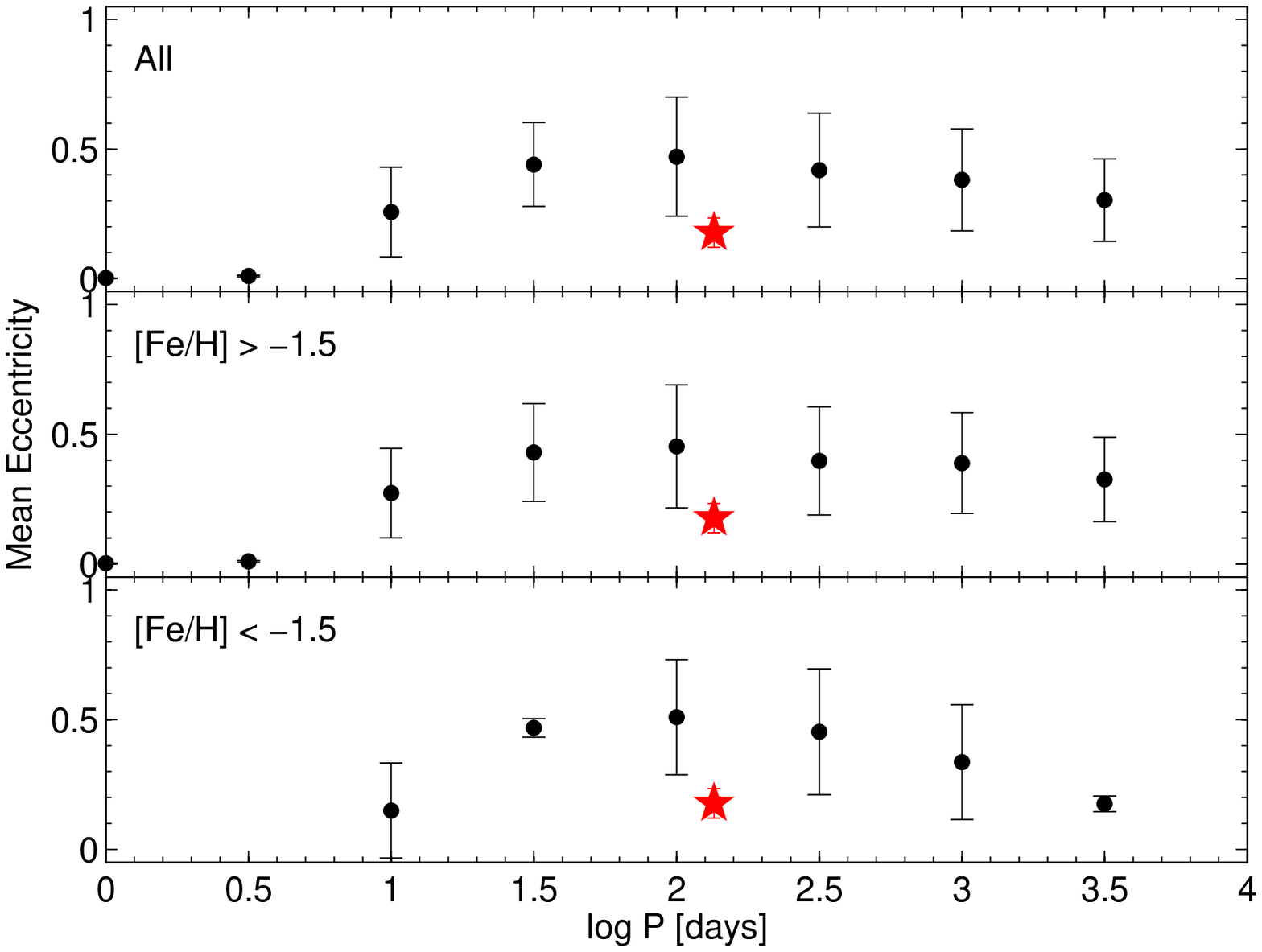}
\end{center}
\caption{ 
Eccentricity-period relation for halo stars from Latham et al. (2002). Metal-rich stars ([Fe/H]$>-1.5$) are shown as open circles, while those below $-1.5$ dex are 
indicated by solid squares. 
 The measurement for Her-3 from this work is illustrated as red star symbol. The lower three panels show the mean relations of the same samples in bins of log\,$P$, where error bars reflect 
 the 1$\sigma$ scatter.}
\end{figure}

At an $e$ of 0.18, Her-3 is clearly not tidally circularized, which is no surprise given its moderate period of 135 d. This can be compared to the 
transition between circularized and eccentric orbits occurring at around 11 days in the Milky Way disk (Duquennoy \& Mayor 1991), at 20 days in halo binaries (Latham 2004), and 
even higher transition periods for metal-poor giants at 70 days (Abt 2006). 
Overall, the fraction of short period binaries is much lower for the metal-poor stars,  although this could in parts be caused by
instrumental and selection effects (Abt 2008).
Aided by Fig.~3, we can conclude that Her-3 is a typical system, consistent with metal-poor giants in the Galactic halo. 

An interesting notion concerns 
the class of carbon-enhanced metal-poor (CEMP) stars, in particular the subclass enhanced in $s$-process elements (CEMP-s). 
In this case, the over-enrichment in these elements is mostly caused by the overflow of 
$s$-rich material synthesized in the AGB-component in a binary system. 
In fact, Lucatello et al. (2005) find that the binary fraction of this kind of metal-poor stars is much higher than $f_b$ in the regular halo field, 
although their periods are probably much shorter on average than in any other Milky Way population. 
This can be contrasted to the occurrence of binarity amongst CEMP-$r$ stars, i.e., those enriched in $r$-process elements: 
Here, Hansen et al. (2011) finds a negligible $f_b$ amongst those stars, which is to be expected since the $r$-process 
proceeds without invoking AGB-nucleosynthesis in any such companion.  

Some of the ultrafaint dSph galaxies are known to host a relatively large number of 
CEMP stars (e.g., Norris et al. 2010; Gilmore et al. 2013) 
and it would be instructive to draw connections to the binarities of these stars and, ideally, on 
$f_b$ of these dSphs. 
However, Her-3 is not suitable for this purpose as it shows neither any significant enhancement in carbon and, on the contrary, 
it is strongly depleted in heavy ($s$- and $r$-process) elements. 
We will return to possible connections with the chemical properties of this star in the next Section.
%
}
\section{Connection with chemical peculiarities{ ?}}
{ Chemical peculiarities are not uncommon in evolved red giants in dSphs
and we can now investigate if binaries stars such as Her-3 can be 
held responsible for some of this galaxy's anomalous abundance patterns. 
}
Koch et al. (2013) found a puzzling deficiency of the
$n$-capture element Ba in all  eleven red giants in their sample of Hercules. 
Furthermore, Her-3  and Her-2 stand out in that
they are depleted in essentially all heavy elements whilst showing
some additional peculiar abundance patterns such as unusually high
Mg/Ca and Co/Cr ratios (Koch et al. 2008). The high Mg/Ca has
later been confirmed by Vargas et al. (2013). 
{ Although Her-2 and Her-3 have peculiar elemental abundances (Koch et al. 2008), 
  the fact that Her-3 is a confirmed binary appears to belie a scenario where the elemental
abundance signatures are coupled to binarity.}
{ Also, the single} red giant Dra~119 in
the Draco dSph galaxy is heavily depleted in $n$-capture
elements. This dSph galaxy is otherwise normal in its Ba-content.

Koch et al. (2008, 2013) argued that the anomalies in the Hercules
dSph galaxy are possibly the imprints of only a few, early massive
($\sim35$ M$_{\odot}$) SNe II.  The low masses of these dwarf galaxies
also incur very low star forming efficiencies, coupled with
stochastic, localized chemical enrichment and inhomogeneous mixing of
the SNe ejecta with the Interstellar Medium (ISM).  However, models of
chemical evolution (e.g., Lanfranchi \& Matteucci 2004; Cescutti et
al. 2006) fail to simultaneously reproduce the observed, low Ba
abundances and trends in, e.g., the $\alpha$-elements with the same,
low star formation prescriptions (Figs.~4,5 in Koch et al. 2013). 

The
identification of binaries in the Hercules dSph can possibly help to illuminate
the question of the chemical peculiarities found in these low-mass
dwarf galaxies.
From the study of the Milky Way stellar halo we know that Ba is,
at early times, produced through the $r$-process (cf. Cescutti et al. 2013). At later times there
is substantial, dominant contribution from the $s$-process in AGB
stars. The onset of $s$-process contribution, at least in the Milky
Way halo, is yet debated. It may occur around [Fe/H]$\sim -2.5$ or as
late as at $-1.4$\,dex (e.g., Simmerer et al. 2004 vs. Roederer et
al. 2010). If Hercules had experienced only the slightest extended 
star formation the contributions from the AGB stars should be
imprinted in the present-day stars.

However, the $r$-process patterns amongst Galactic halo stars
can vary broadly from star to star, arguing for an inhomogenous
injection mechanism into the early ISM. This is in contrast to the
fairly uniform (and low) Ba-abundance in the Hercules dSph galaxy.
This could indicate that some of the sources for the $r-$process may
not be operational in Hercules, that mixing in this galaxy was highly
inefficient, or that only a certain mass range for contributing SNe is
favoured; the reasons for either scenario are yet unknown.

Can binaries help explain these abundance patterns? 
{ Hansen et al. (2011) }
argued that binary stars, at least in the Milky Way halo,
played no special role in producing $r$-process elements in the early
Galaxy, while they are clearly an important source of later
$s$-process elements (e.g., K\"appeler et al. 2011). Here it is
noteworthy that the CMD for the Hercules dSph galaxy (Brown et al. 2012) 
indicate a substantial population of blue
stragglers, which implies a binary fraction, $f_b$, as large as
35--60\% (T. Brown, priv. comm.).

Binary evolution can be very efficient at inhibiting the return of
$s$-process elements from the system's AGB component to the ISM, 
{ or can even fully prevent the original donor star from going 
through its AGB phase (McCrea 1964)}.
Depending on the binary separation the component that would evolve onto the
AGB first will fill its Roche lobe.  During the common-envelope phase,
the star's outer layers will be lost, preventing the thermal pulses to
occur, switching off any production of Ba through the $s$-process
(Izzard 2004; Izzard et al. 2006).  From a theoretical point of view,
it is known that AGB stars at low metallicities do not undergo the
third dredge up, thereby inhibiting the $s$-process to occur (Lau et
al. 2008, 2009).

{ In terms of star formation in the Hercules dSph, 
it cannot be excluded that 
this galaxy may indeed have had an extended
star formation history.
} 
There are two lines of evidence for this:
a broadened turn-off in the CMD and a substantial spread in iron abundances and the Ca/Fe ratios (Ad\'en et al. 2011).

Brown et al. (2012) show that the well resolved turn-off region in the
CMD of Hercules has sufficient spread that a small
contribution from a population younger by 1--2 Gyr cannot be { ruled out}.
This presumes that this component, of at most 10\% (by number),
resides at the higher metallicity tail of the metallicity
distribution.  Other, recent observations of the star formation
history of Hercules suggest that $80\pm10$\% of its total stellar mass
formed prior to 10 Gyr ago, while the remainder of star formation
truncated around $\sim$8 Gyr ago (D. Weisz, priv. comm.).  This adds
to the question of why Ba is so low and seemingly homogeneous, whilst
the galaxy has had an inkling of an extended star formation history
and experienced low-level chemical evolution.
  
The presence of a substantial iron-spread in Hercules indicates that
SNe Ia have participated in the chemical enrichment. Given its low
star formation rate and the very metal-poor location of the turn-over
in the [$\alpha$/Fe] distribution, those SN must have occurred at
early times, rendering prompt SNe Ia a viable option for
this enrichment -- these could have contributed as early as within 50
Myr after the formation of their progenitors (e.g., Mannucci et
al. 2006; Matteucci et al. 2009).  A fraction of those progenitors
 can be expected to have undergone the AGB phase before
losing their envelope in a wind, finally exposing the white dwarf that
partakes in the SN explosion.

However, to explain the elemental abundance patterns in Hercules, one
would need to shut off almost all Ba-production in those early AGB
stars, in turn implying that the majority of binaries in Hercules need
to be in close and short-period systems so that Roche-lobe overflow
can take place.  
Unless the distribution of orbital
elements in the dSph galaxies is significantly different from that in
the Solar neighbourhood and the halo (Duquennoy \& Mayor 1991), this
is a very unlikely scenario, and Minor (2013)  suggests that they are in fact comparable. 
 Furthermore, we measured the H$\alpha$
bisector velocities of the red giants Her-2 and Her-3, which, at 1--5
km\,s$^{-1}$, are very slow and thus indicate that these stars
currently do not experience any active overflow.

Moreover, the above scenario alone cannot explain the origin of other
abundance anomalies such as the high Mg/Ca ratios so that one still
needs to invoke the massive stars in Hercules' enrichment
history. These would have imprinted their signatures in the gas out of
which the future generations of binary progenitors have formed.

If the secondary star to Her-3 was presently on the main sequence, as
cannot be ruled out from our orbital solution, it would still be
unevolved as yet without any influence on the chemical budget of the
binary system.
This would imply that, in Her-3, we indeed observe the unaltered,
primordial material out of which the stars were born, which is in line
with the particular Mg/Ca and Co/Cr ratios observed in those stars
(e.g., Heger \& Woosley 2010; Nomoto et al. 2010).
%
%
\section{Summary and discussion}
Based on extensive radial velocity measurements of a red giant in the faint Hercules dSph, we could identify it with 
a moderately eccentric, close binary in a 135-days orbit. Detailed chemical element abundance measurements in 
past studies have revealed that the red giant primary shows a number of peculiar abundance patterns that are, 
to a similar extent, also found in other stars of this galaxy (Koch et al. 2008, 2013; Ad\'en et al. 2011; Vargas et al. 2013). 
A broad metallicity range of more than one dex in iron abundance and some of the abundance anomalies suggest that 
Hercules has experienced very inefficient star formation and was probably governed by the explosion of only a few, massive SNe. 
The characterization of Her-3 as a binary allowed us to take another angle on the abundance peculiarities in this work, although 
the required construct of many close binaries suggests that binaries were likely not the main reason for the peculiar chemical enrichment this galaxy experienced.  

In order to assess whether (close) binaries play 
a significant role for the chemical evolution of dSphs through the mechanisms addressed in Sect.~6 
 it would be important to
correlate the abundance of Ba and other $s$-process elements with the
binary fraction in different dSphs. Unfortunately, estimates of $f_b$
are still sparse for these galaxies:  CMDs and kinematic data of the
remote Leo~I and II dSphs place upper limits of $\sim40$--60\%
(Gallart et al. 1999; Koch et al. 2007a,b), but Shetrone et al. (2003)
found slightly supersolar [Ba/Fe] ratios in two stars in these galaxies at moderately high
metallicities.
Typical fractions in other dSphs studied to date lie around 20--50\%
(Scl; Queloz et al. 1995); 20--30\% (UMi and Dra\footnote{Note,
  however, that the {\em dynamically} significant $f_b$ is generally
  much smaller, e.g., 5\% in the case of Dra (Kleyna et al. 2002).};
Olszewski et al. 1996), and $\sim40$\% (For; Walker et al. 2006).  No
study has been carried out in any of the UFDs so far, and nothing is known 
about the distributions of orbital parameters, in particular their separations. 
Thus the Ba-binary connection  can to date only rest on on a comparison of the Ba-poor Hercules dSph with
its possibly high binary fraction and the benchmark $n$-capture depleted star Dra~119 (Fulbright et al. 2004), a member of  
an otherwise (Ba-) normal environment, namely the Draco dSph, 
at a lower $f_b$ of maximally 30\%. 

To fully disentangle the Her-3-system's individual components, thus gaining a better insight in their mass-ranges and evolutionary 
status, it would be necessary to obtain highly precise light curves to look for optical variability (Musella et al. 2012). 
\acknowledgments
AK acknowledges the Deutsche Forschungsgemeinschaft for funding from  Emmy-Noether grant  Ko 4161/1. 
This work was in part supported by Sonderforschungsbereich SFB 881 "The Milky Way System" (subproject A4) of the German Research Foundation (DFG).
MIW acknowledges the Royal Society for a support through a University Research Fellowship.
We thank I. Ivans and J. Marshall for obtaining parts of the observations, R. Izzard, J.-C. Passy, and I.B. Thompson for helpful discussions, and 
I. Musella for providing photometry for this star.
\end{document}